\documentclass[final,5p,times,twocolumn]{elsarticle}

\usepackage[utf8x]{inputenc}
\usepackage{amssymb}
\usepackage{amsmath}
\usepackage{amsthm}
\usepackage{amsmath}
\usepackage{hyperref}
\usepackage{subfigure}
\usepackage[percent]{overpic} 
\usepackage{picture}
\usepackage{color}
\usepackage{nth}
\usepackage{gensymb}
\usepackage{lineno} 
\usepackage{textcomp}
\usepackage{threeparttable}

\graphicspath{{img/}}

\journal{Nuclear Instruments and Methods in Physics Research A }

\begin{document}

\begin{frontmatter}


\title{Recent results at SPARC\_LAB}


\author[lnf]{R. Pompili}\ead{riccardo.pompili@lnf.infn.it}
\author[lnf]{M.P. Anania}
\author[lnf]{M. Bellaveglia}
\author[lnf]{A. Biagioni}
\author[lnf]{S. Bini}
\author[lnf]{F. Bisesto}
\author[lnf]{E. Chiadroni}
\author[roma2]{A. Cianchi}
\author[lnf]{G. Costa}
\author[lnf]{D. Di Giovenale}
\author[lnf]{M. Ferrario}
\author[lnf]{F. Filippi}
\author[lnf]{A. Gallo}
\author[lnf]{A. Giribono}
\author[lnf]{V. Lollo}
\author[lnf]{A. Marocchino}
\author[lnf]{V. Martinelli}
\author[roma1]{A. Mostacci}
\author[lnf]{G. Di Pirro}
\author[lnf]{S. Romeo}
\author[lnf]{J. Scifo}
\author[lnf]{V. Shpakov}
\author[lnf]{C. Vaccarezza}
\author[lnf]{F. Villa}
\author[racah]{A. Zigler}

\address[lnf]{Laboratori Nazionali di Frascati, Via Enrico Fermi 40, 00044 Frascati (Rome), Italy}
\address[roma2]{University of Rome "Tor Vergata" and INFN-Roma Tor Vergata, Via della Ricerca Scientifica 1, 00133 Rome, Italy}
\address[roma1]{University of Rome "Sapienza", Piazzale Aldo Moro 5, 00185 Rome, Italy}
\address[racah]{Racah Institute of Physics, Hebrew University, 91904 Jerusalem, Israel}

\begin{abstract}
The current activity of the SPARC\_LAB test-facility is focused on the realization of plasma-based acceleration experiments with the aim to provide accelerating field of the order of several GV/m while maintaining the overall quality (in terms of energy spread and emittance) of the accelerated electron bunch.
In the following, the current status of such an activity is presented. We also show results related to the usability of plasmas as focusing lenses in view of a complete plasma-based focusing and accelerating system.
\end{abstract}


\end{frontmatter}


\section{Introduction}\label{intro}
Plasma-based acceleration, either driven by ultra-short laser pulses or electron bunches, represents one of the most promising techniques able to overcome the limits of conventional RF technology and allow the development of compact facilities.
State of the art accelerators are indeed based on RF technology that allows to reach maximum fields of the order of 150~MV/m~\cite{wu2017high}. Such a limit, mainly due to electrical breakdown, requires therefore the development of km-size or even larger facilities if the energy has to be further increased~\cite{evans2008lhc,oide2016design}.
By enhancing the accelerating gradient the overall compactness can be improved and costs reduced.
So far several techniques based on dielectric~\cite{o2016observation} and plasma~\cite{1979PhRvL..43..267T} wakefield acceleration have demonstrated that huge fields of the order of few tens of GV/m can be produced by a \textit{driver} pulse (a laser or a particle bunch) and efficiently used to accelerate a \textit{witness} bunch in cm-scale structures~\cite{geddes2004high,leemans2006gev,faure2006controlled}.
As a particle beam-driven scheme, Plasma Wakefield Acceleration (PWFA)~\cite{2007Natur.445..741B,litos2014high} represents one of the best candidates among the other plasma-based methods. By means of high-energy and high density driver bunches, PWFA can provide extremely high accelerating fields for very long distances~\cite{lu2005limits,lu2009high} if compared to laser-driven schemes, mainly limited by non-negligible effects like diffraction, dephasing and depletion of the laser pulse while propagating through a plasma~\cite{1996ITPS...24..252E,2009RvMP...81.1229E}.

In the following we discuss the current status of the PWFA experiment planned at the SPARC\_LAB test-facility~\cite{ferrario2013sparc_lab}. Here several particle beam-driven configurations will be tested (consisting of one or more driver bunches with different charge profiles) with the goal to achieve large accelerating fields (of the order of several GV/m) while preserving the overall quality (in terms of energy spread and emittance) of the accelerated electron bunch. In this regard we focus on the development and experimental test of all the equipment necessary to perform such an experiment, including its diagnostics.
We also show some preliminary results obtained by using the plasma itself as a focusing lens.

\section{Plasma Wakefield acceleration at SPARC\_LAB}\label{pwfa_sparc}
There are several possible PWFA scenarios when operating such experiments, mainly depending on the working regime (linear or nonlinear). The \textit{linear} regime has the advantage that the plasma oscillation may be resonantly driven, but has the notable drawback that the focusing is nonlinear in radial coordinate \textit{r} and the same dependence would apply for the provided acceleration. In the \textit{blow-out} regime~\cite{rosenzweig1991acceleration} the beam density is much greater than the plasma density ($n_b\gg n_p$) and the beam channel is completely rarefied of plasma electrons. Under these conditions the plasma response is very nonlinear, the plasma wavelength $\lambda_p$ depends on the wakefield amplitude and large wave-breaking spikes are excited at the end of the first plasma oscillation.
To overcome these limitations, the beam-driven experiment at SPARC\_LAB will operate in the so-called quasi-nonlinear (QNL) regime~\cite{rosenzweig2010plasma,rosenzweig2012plasma}. We can define a dimensionless charge $\widetilde{Q} \equiv {N_b k_p^3}/{n_p} = 4 \pi k_p r_e N_b$, with $N_b$ the driver number of particles, $k_p = 2\pi / \lambda_p$ and $r_e$ the classical electron radius, that quantifies if the regime is linear ($\widetilde{Q} \ll 1$) or nonlinear ($\widetilde{Q} > 1$).
A QNL regime is given by the condition $\widetilde{Q}<1$ for linearity combined with $n_b / n_p >1$ for bubble formation, with $n_b$ the bunch charge density. By indicating with $E_{WB}~\left[V/m\right] \approx 96 \sqrt{n_p (cm^{-3})}$ the wave-breaking field, the accelerating field is approximately given by~\cite{lu2005limits}
\begin{equation}
E_z \approx 1.3 E_{WB} \frac{n_b}{n_p} k_p^2 \sigma_r^2 \log \left({\frac{1}{k_p \sigma_r}}\right)~.
\end{equation}
Since $\lambda_p$ is not amplitude-dependent, the QNL regime can be used also in resonant schemes with multiple drivers separated in time by $\lambda_p$~\cite{kallos2008high} allowing to produce an accelerating field proportional to the number of drivers while handling bunches with lower charges and emittances.


\section{Experimental apparatus}\label{exp_apparatus}
\begin{figure}[h]
\centering
\includegraphics[width=0.7\linewidth]{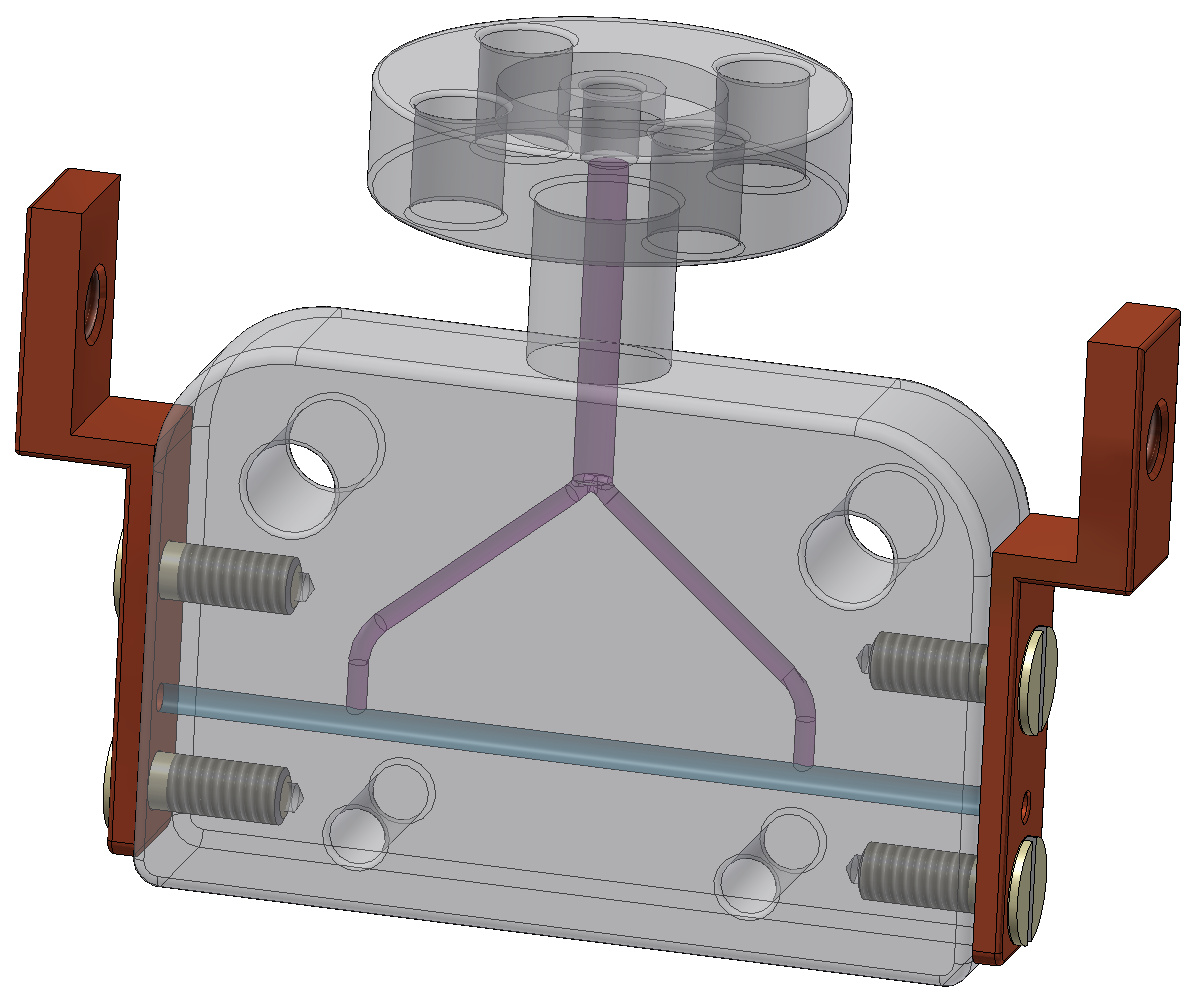}
\caption{Setup of a 3~cm-long capillary with its injection system. Hydrogen gas fills the entire volume through two symmetric inlets located at $1/4$ and $3/4$ of its length. The capillary diameter is 1~mm. The two electrodes are directly connected to the HV discharge circuit to ionize the gas and produce plasma.}
\label{capillary}
\end{figure}

At SPARC\_LAB, the plasma is produced by ionizing Hydrogen gas contained in capillary tubes (1-10~cm lengths) made either from sapphire or plastic material. A HV discharge circuit (up to 20~kV voltage between the electrodes and 240~A peak current) is used for this purpose~\cite{anania2016plasma,filippi2016spectroscopic}. The experimental setup of a 3~cm-long capillary is shown in Fig.~\ref{capillary}. The PWFA experiment will operate with plasma density of the order of $10^{16}$~cm$^{-3}$.
In the following, the main systems (beam diagnostics, injection and extraction) involved into the PWFA experiment will be presented.

\subsection{Beam diagnostics}\label{beam_diag}
To properly perform and take control of the entire plasma acceleration process, several diagnostics have to be implemented. The diagnostics must be single-shot both for the transverse and longitudinal dimensions in order to correlate the properties of the injected beams with the accelerated ones.
The longitudinal profile of the input beam has to be carefully characterized. This is particularly important if resonant schemes employing multiple driver bunches are adopted: in such a case the distance between the drivers determines the resulting accelerating field felt by the witness~\cite{pompili2014first}. Also its transverse spot at the plasma entrance plays a key role since its Twiss beta-function must be matched with the plasma channel~\cite{barov1994propagation,pompili2016beam}.
In the next sections the main diagnostic devices related to the PWFA experiment are described.

\subsubsection{Electro-Optical Sampling}
The EOS is a non-intercepting and single-shot device that allows to monitor the bunch longitudinal profile and time of arrival~\cite{wilke2002single}. At SPARC\_LAB we employ $100~\mu m$-thick Zinc Telluride (ZnTe) and Gallium Phosphide (GaP) crystals with a Ti:Sa laser ($\lambda\approx 800$~nm, $\sigma_t\approx 80$~fs rms) as probe, resulting in a temporal resolution on the bunch duration down to $\sigma_t \approx 90$~fs (rms).
The probe is directly split from the photo-cathode laser system, resulting in a natural synchronization with the electron beam. This solution allows also to directly measure the relative timing-jitter between the photo-cathode laser and the beam, a fundamental parameter in view of laser-driven plasma acceleration through the external injection of a linac-accelerated witness bunch~\cite{eos_jitter}.
The encoding of the beam longitudinal profile is then obtained with the spatial decoding setup~\cite{cavalieri2005clocking}, in which the laser crosses the nonlinear crystal at an angle of $30\degree$.

\begin{figure}[h]
\centering
\subfigure{
\begin{overpic}[width=0.8\linewidth]{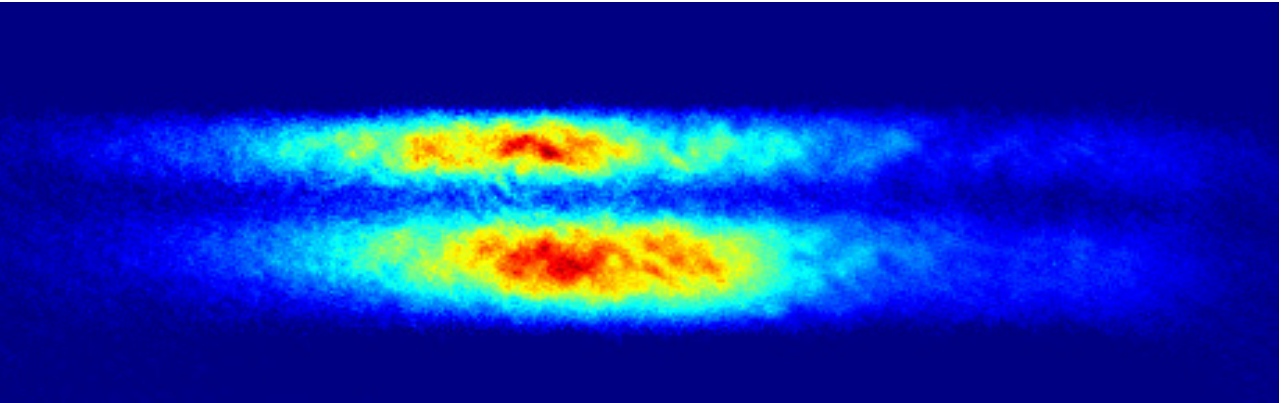}
\put(5,25){\color{white}(a)}
\end{overpic}
\label{comb_gap}
}
\subfigure{
\begin{overpic}[width=0.9\linewidth]{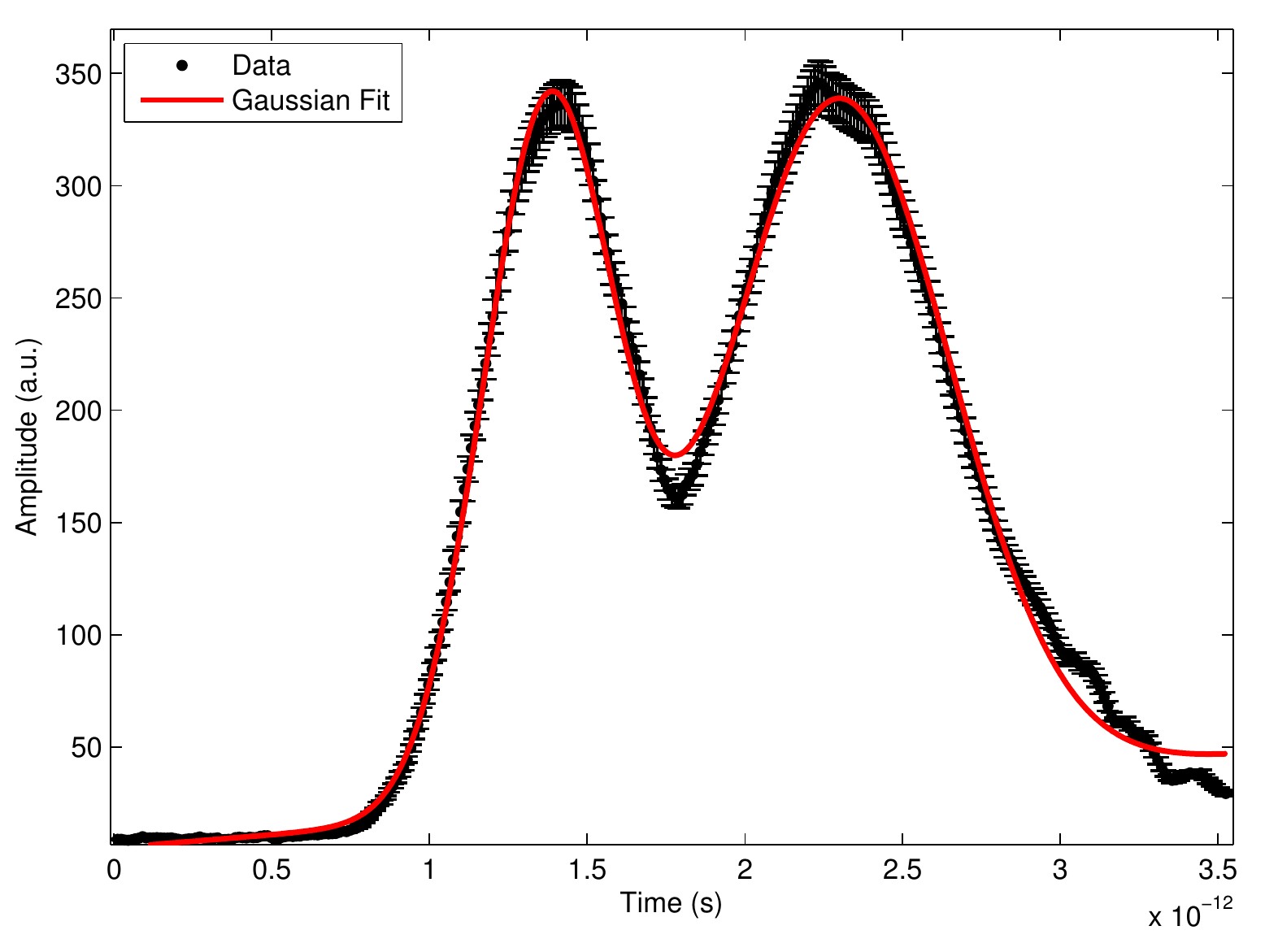}
\put(10,60){\color{black}(b)}
\end{overpic}
\label{comb_gap_prof}
}
\caption{(a) EOS signals resulting from a comb-like beam consisting of two 80~pC bunches. (b) Resulting temporal profile. The retrieved bunch durations are 160~fs and 200~fs (rms). The bunch distance is 800~fs.}
\label{eos_signals}
\end{figure}

Figure~\ref{eos_signals} show a measured electro-optic signal produced by a $100~\mu m$-thick GaP crystal located close to a 110~MeV comb-like train consisting of two bunches with 80~pC charge each one separated by 800~fs. Such a train configuration is directly realized on the photo-cathode by means of the laser-comb technique~\cite{2011NIMPA.637S..43F}. Since the EOS is a non-intercepting and single-shot device, such results confirm that it represents a valuable tool in order to monitor the longitudinal profile of the beam before its injection into the plasma.

\subsubsection{Transition Radiation}
Transition radiation (TR) can be used both as longitudinal and transverse beam diagnostics. It is produced when the electron bunch crosses an aluminum-coated silicon screen (in our case it is oriented at $45\degree$ with respect to the beam line).
In order to reconstruct the longitudinal profile of the plasma-accelerated bunches we use a Michelson interferometer measuring the coherent part of TR (CTR)~\cite{giorgianni2016strong,giorgianni2016tailoring}.
It consists of two highly polished mirrors and a $12~\mu m$-thick mylar layer acting as a beam-splitter. Only the backward radiation is collected. It is extracted through a diamond window and collimated by a $90\degree$
off-axis parabolic mirror towards a flat mirror reflecting the radiation to the interferometer. The light is then split in two beams. One is transmitted towards a fixed mirror while the other is reflected in the direction of a movable one. The beams are then recombined on the beam-splitter and are measured by pyro-electric detectors (0.5–30 THz spectral range). The radiation is collected by the detector at several positions of the movable mirror, producing an interference pattern used to reconstruct the beam temporal profile with a resolution $\sigma_t\approx 20$~fs (rms)~\cite{eos_jitter}.

\begin{figure}[h]
\centering
\includegraphics[width=1.0\linewidth]{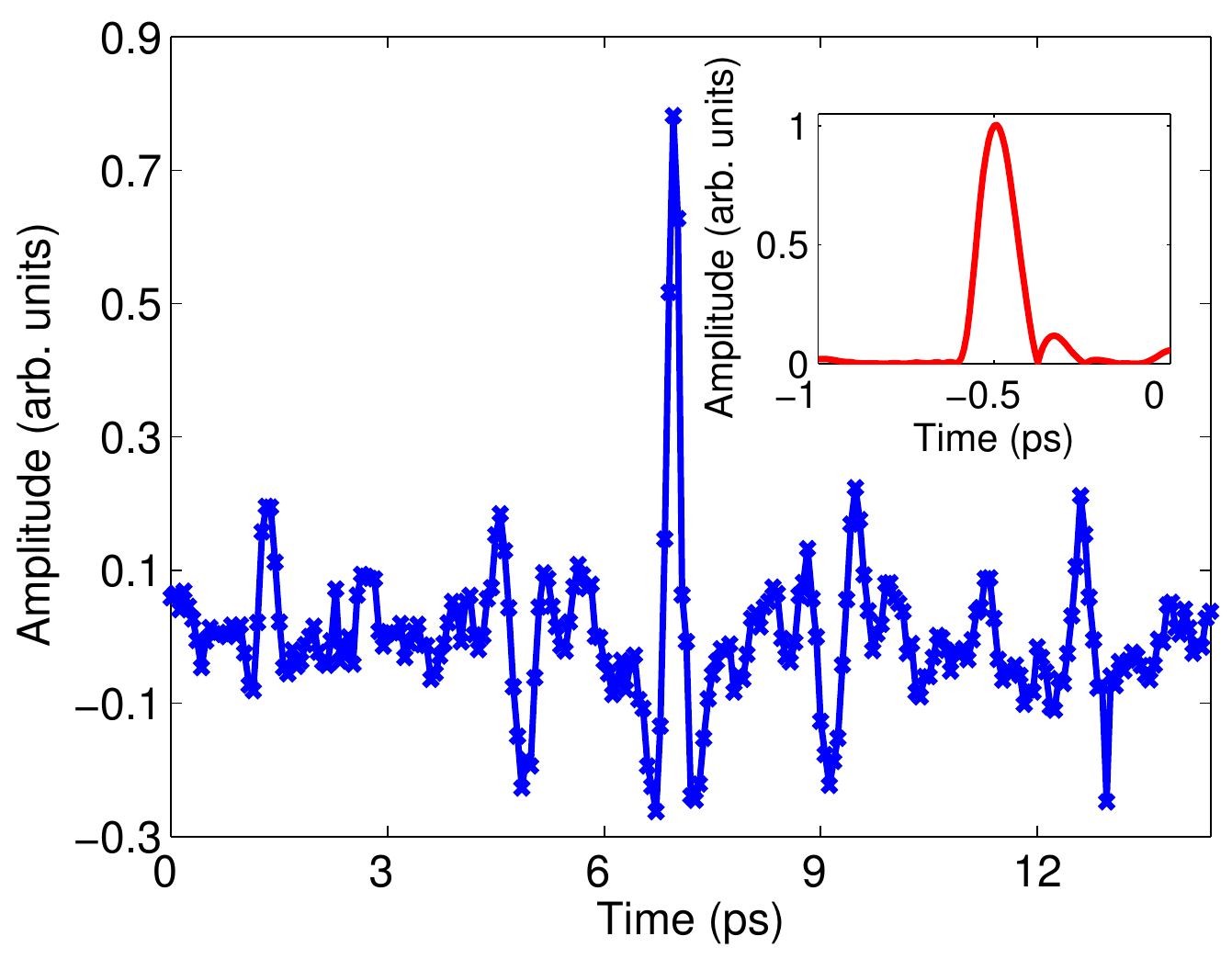}
\caption{Interferogram obtained with Michelson interferometer. The inset shows the retrieved bunch temporal profile. The
resulting in bunch duration is $\sigma_t\approx 86$~fs (rms).}
\label{BunchInterf}
\end{figure}

Figure~\ref{BunchInterf} shows a typical raw interferogram. Data have been acquired by changing the position of the interferometer movable mirror with $10~\mu m$ steps. In order to reconstruct the bunch temporal profile from the interferogram, the effect of the finite-size of the CTR screen must be considered~\cite{castellano1999effects}. The electron bunch profile is then retrieved by means of the Kramers–Kronig relations~\cite{giorgianni2015intense}. The reconstructed bunch profile is shown in the inset of Fig.~\ref{BunchInterf}. The retrieved duration is $\sigma_t\approx 86$~fs (rms).

\subsection{Plasma diagnostics}\label{plasma_diag}
\begin{figure}[h]
\centering
\includegraphics[width=1.0\linewidth]{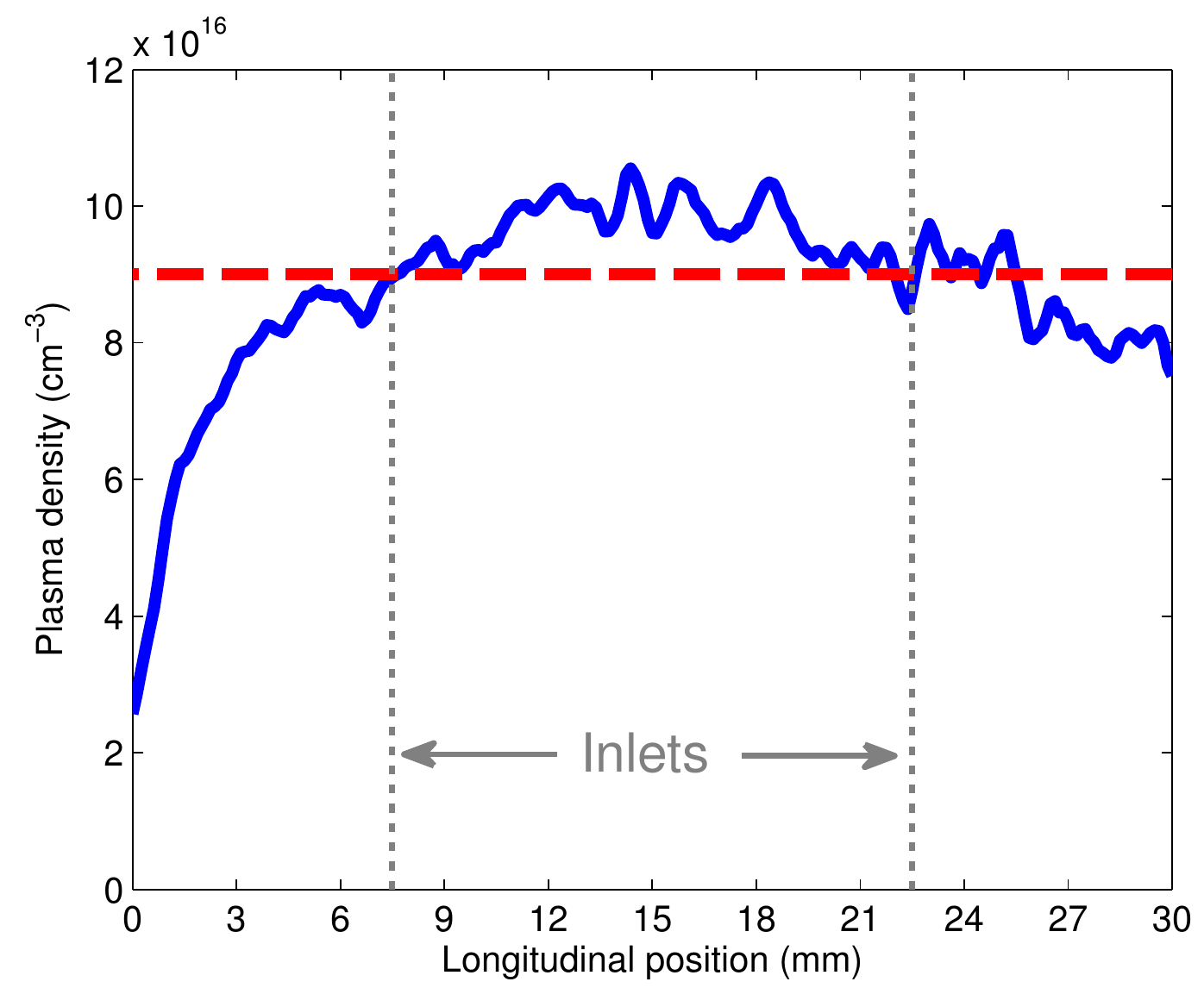}
\caption{Longitudinal profile of the plasma density measured with Stark broadening-based diagnostics. Each point is obtained by averaging 5 single-shot measurements. The resulting error is of the order of $10\%$~\cite{filippi2016spectroscopic}. The mean value (red line) is $n_p\approx 9\times 10^{16}$~cm$^{−3}$. The gray dashed lines refer to the position of the two capillary inlets.}
\label{StarkDensity}
\end{figure}

Hydrogen atoms, when excited, emit light at a different wavelengths in the visible range and are usually referred as Balmer lines. Their linewidth is caused by several effects. The main one is represented by Stark broadening and is due to the presence of an external electric field (e.g. nearby free electrons) acting on the emitter. By measuring the linewidth we can therefore reconstruct the electron density of the plasma near the emitter. For hydrogen plasma, the full width at half maximum (FWHM) of the Balmer alpha line $\Delta \lambda$ is related to the plasma~\cite{filippi2016spectroscopic} as $\Delta \lambda \left[{nm}\right]=\alpha(n_e,T) n_e \left[10^{18}~cm^{-3}\right]^{2/3}$.
Figure~\ref{StarkDensity} shows the longitudinal plasma density profile measured along the 3~cm-long sapphire capillary. The density was retrieved from Stark broadening-based measurements of the $H_{\beta}$ Balmer line. The light was collected by a lens and then analyzed through a CCD spectrometer, where the $H_{\beta}$ linewidth is quantified.

\subsection{Beam injection and extraction}\label{PMQ_apparatus}
The reduction of the bunch transverse dimensions requires a proper focusing system~\cite{pompili2016beam}.
Such a task can not be accomplished with normal-conducting electromagnetic solenoids or quadrupoles because the provided field gradients would be too weak. 
In this context the state of the art is represented by permanent-magnet quadrupoles~\cite{lim2005adjustable,becker2009characterization,nichols2014analysis} (PMQ).
So far field gradients of about 560~T/m have been reported~\cite{lim2005adjustable} when using small-bore PMQs manufactured with neodymium-iron-boride (NdFeB) materials.

\begin{figure}[h]
\centering
\subfigure{
\begin{overpic}[height=0.23\textwidth]{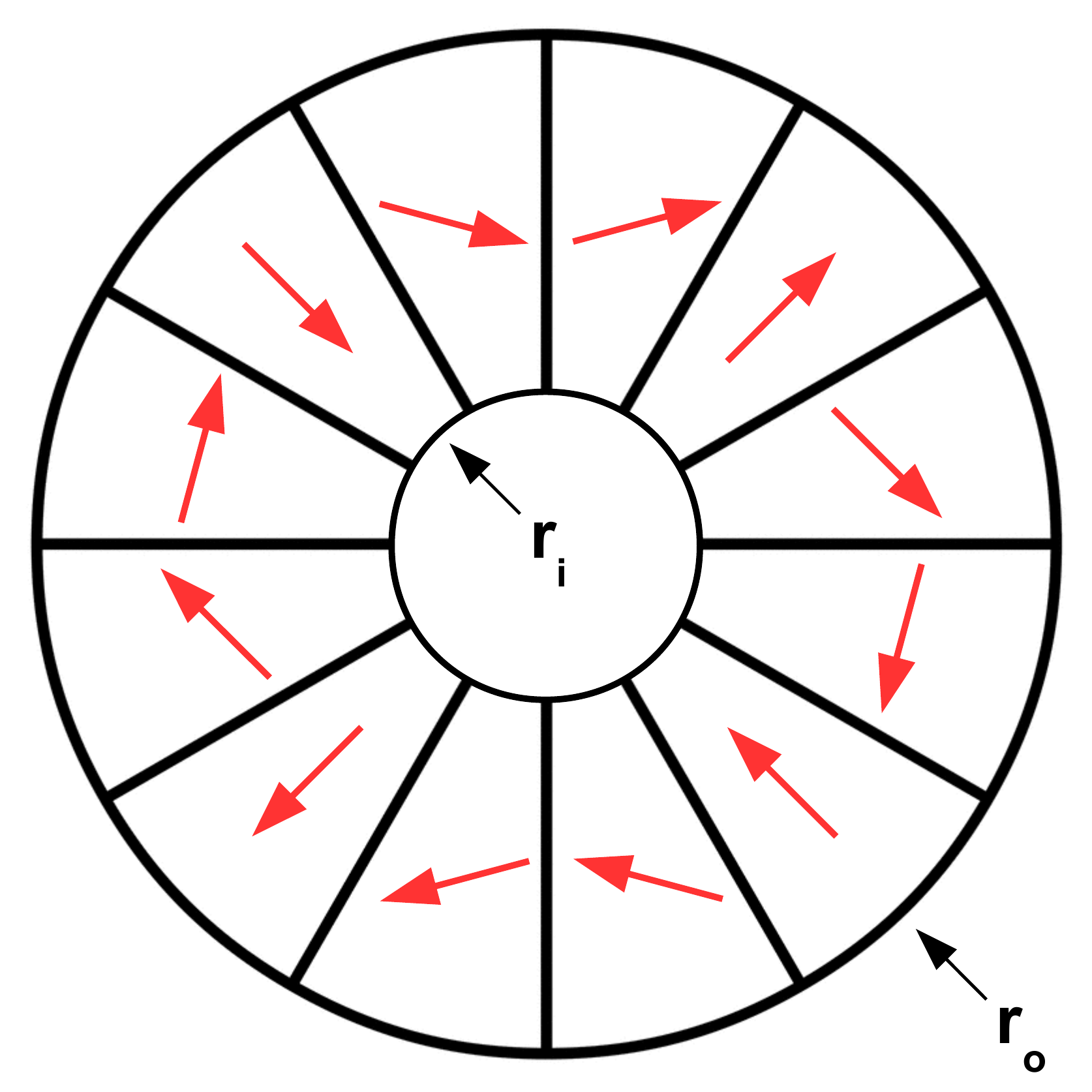}
\put(2,93){\color{black}\textbf{a}}
\end{overpic}
\label{PMQside}
}
\subfigure{
\begin{overpic}[height=0.23\textwidth]{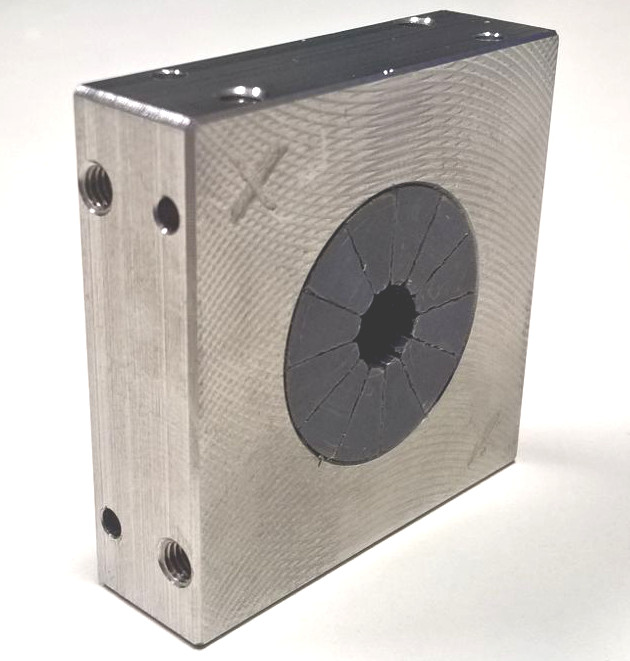}
\put(2,93){\color{black}\textbf{b}}
\end{overpic}
\label{PMQpict}
}
\caption{(a) Sketch of a single PMQ with 12 sectors. The magnetizations are oriented as the red arrows. To produce the required quadrupolar field configuration, the magnetization is rotated by $60\degree$ after each sector. (b) PMQ manufactured by KYMA ($r_i=3$~mm, $r_o=10$~mm). The cross on the upper left side indicates the sector whose magnetic field is pointing inwards.}
\label{PMQpics}
\end{figure}

Quadrupoles based on permanent magnets have been developed by following the so-called Halback geometry~\cite{halbach1980design}, that consists of an array of $M$ uniformly magnetized segments with trapezoidal shape arranged into a ring with an inner (outer) radius $r_i$ ($r_o$). It is shown in Fig.~\ref{PMQside}, where the number of segments is considered to be $M=12$. To guarantee the generation of a quadrupolar field, the magnetization between two consecutive sectors must rotate by $4\pi/M$.
Based on the results obtained with numerical simulations about the optimal configuration required for the PWFA experiment, we have adopted a mechanical design consisting of $M=12$ sectors with $B_r=1.32$~T, $r_i=3$~mm and $r_o=10$~mm. Figure~\ref{PMQpict} shows a sample of a PMQ realized according to this layout. The resulting field gradient $G \approx 520$~T/m. The PMQs will be employed in two identical triplet systems both for the injection and extraction from the plasma.

\subsection{Vacuum chamber}\label{vacuum_chamber}
All the equipment necessary to perform the PWFA experiment is hosted in a ultra-high vacuum (UHV) compatible experimental chamber. Many efforts have been made on its realization since the Hydrogen-filled capillary is very close to the RF linac (less than a meter) and any contamination of external gas in such UHV environment has be avoided. For this purposes we have realized a differential system consisting of several pumping stages.

\begin{figure}[h]
\centering
\includegraphics[width=0.95\linewidth]{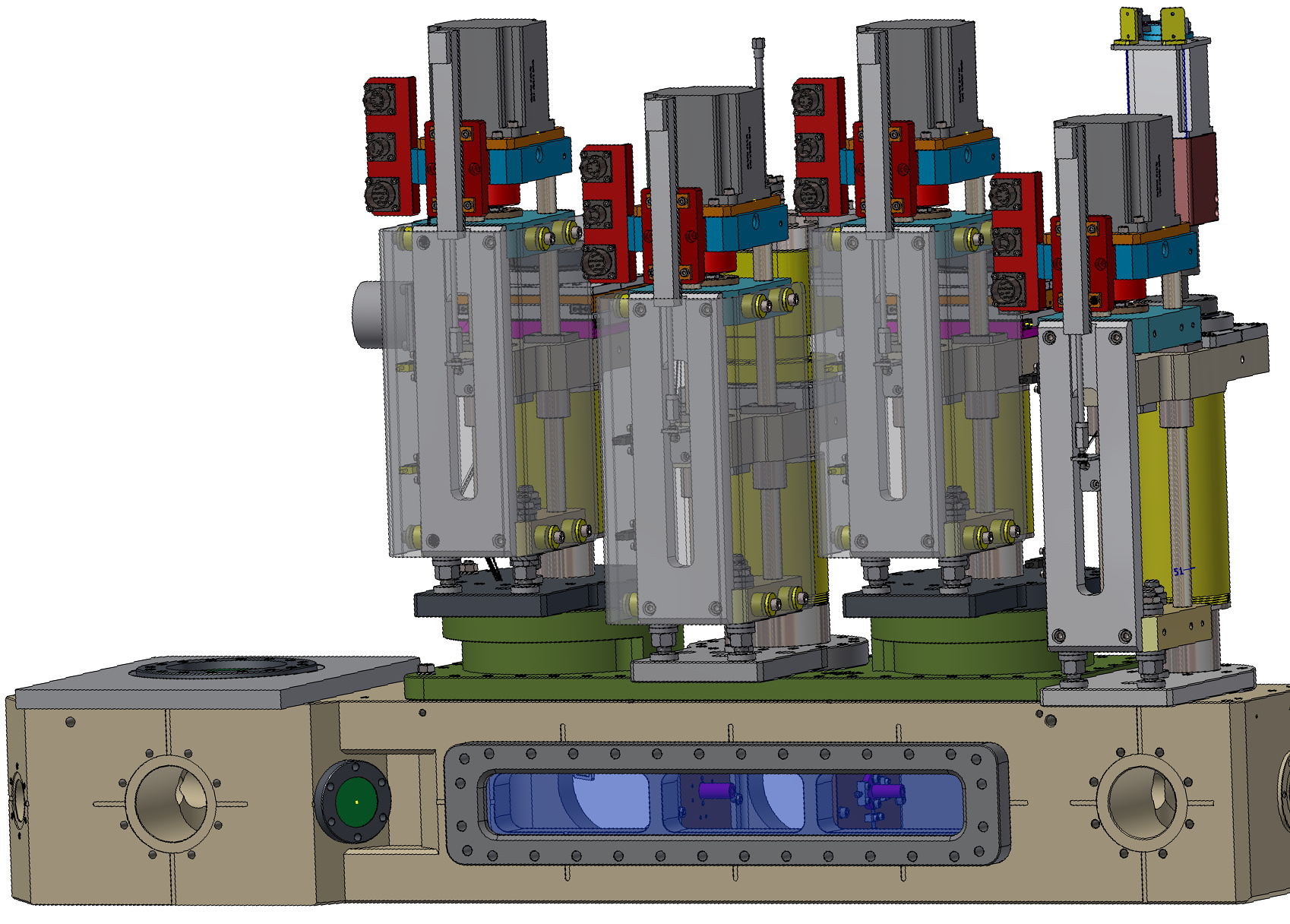}
\caption{Vacuum chamber for the PWFA experiment. The beam enters from left. The first station is for transverse diagnostics and EOS. The large middle one hosts the capillary, its injection system and two symmetric triplets of PMQs to focus in and extract the beam from the capillary. The last station is for transverse diagnostics and THz radiation generation.}
\label{NewCombChamber}
\end{figure}

The layout of the vacuum chamber is shown in Fig.~\ref{NewCombChamber}. The electron beam coming from the SPARC\_LAB linac enters from the left. We have, in cascade, the EOS diagnostics, the input triplet of PMQs for the beam injection, the capillary for the PWFA, the exit triplet for the beam extraction and the TR diagnostics.
The entire system is mounted on a three axis motorized stage allowing to horizontally translate it and tune its horizontal and vertical tilts. As shown in Fig.~\ref{NewCombChamber}, the vertical position of the PMQs, capillary and TR screen can be finely adjusted in order to be centered along the beam path.
The Hydrogen gas is injected into the chamber through a fast solenoid valve with 3~ms opening time.
To confine the flowing gas to the central part of the vacuum chamber, several vacuum impendences (6~mm diameter) have been realized and installed in order to dissect all these sub-chambers. In such a way we can safely have up to 20~mbar of Hydrogen gas flowing at 1~Hz into the capillary while keeping the vacuum at the entrance and exit of the entire vacuum chamber below $\approx 5\times 10^{-8}$~mbar.

\begin{figure}[!h]
\centering
\subfigure{
\begin{overpic}[width=0.45\textwidth]{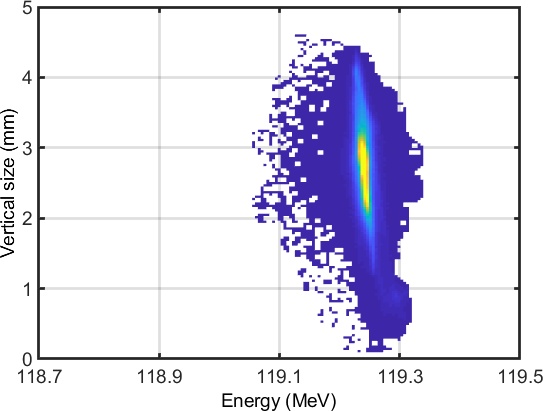}
\put(10,67){\color{black}\textbf{a}}
\put(73,67){\color{red}\textbf{Plasma OFF}}
\end{overpic}
\label{energy_noplasma}
}
\subfigure{
\begin{overpic}[width=0.45\textwidth]{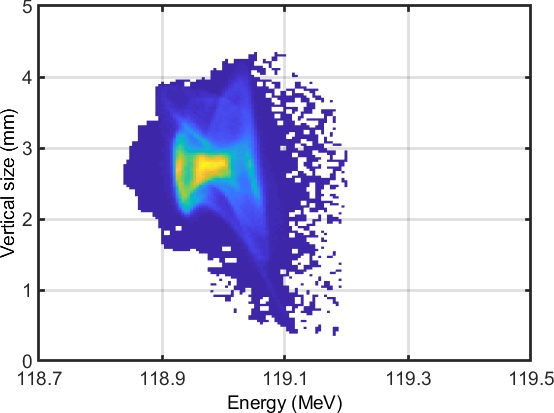}
\put(10,67){\color{black}\textbf{b}}
\put(73,67){\color{red}\textbf{Plasma ON}}
\end{overpic}
\label{energy_withplasma}
}
\caption{Electron bunch energy spectrum with plasma off (a) and on (b). In (a) the central energy is 119.2~MeV (25~keV energy spread). In (b) the central energy is 119.0~MeV (47~keV energy spread). These snapshots are obtained on a screen located downstream a magnetic spectrometer.}
\label{energy_plasma}
\end{figure}
%

\section{Experimental results}\label{exp_results}
From the beginning of 2016, some preliminary results have been obtained. Figure~\ref{energy_noplasma} shows the energy profile of a 50~pC beam (1~ps rms duration, $1~\mu m$ emittance) when passing through the capillary without any plasma. With the first PMQ triplet its transverse sizes have been focused down to $\sigma_{x,y}\approx 30~\mu m$.

\begin{figure*}[!h]
\centering
\subfigure{
\begin{overpic}[width=0.3\textwidth]{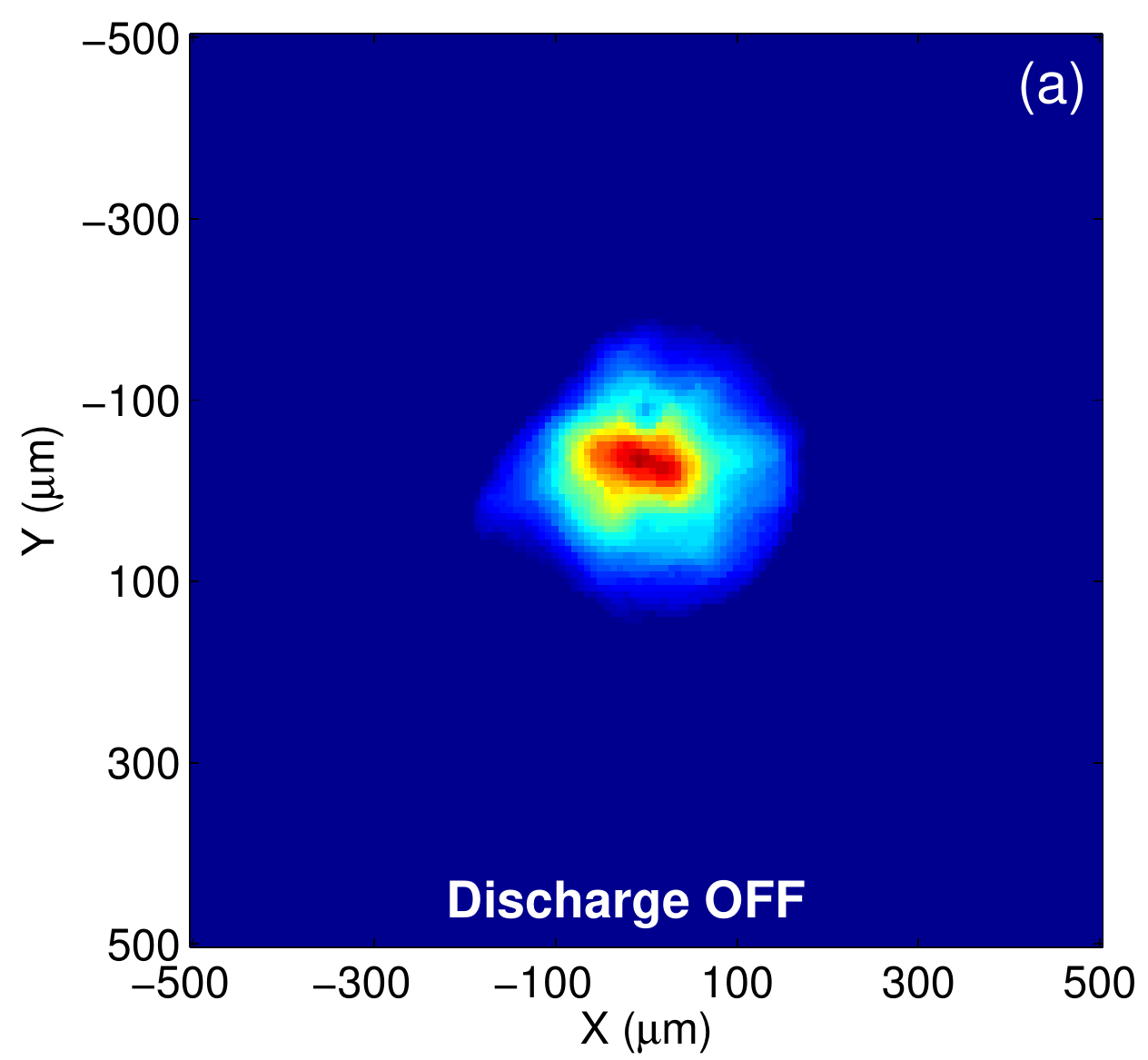}
\put(20,80){\color{white}\textbf{a}}
\end{overpic}
\label{SpotOFF}
}
\subfigure{
\begin{overpic}[width=0.3\textwidth]{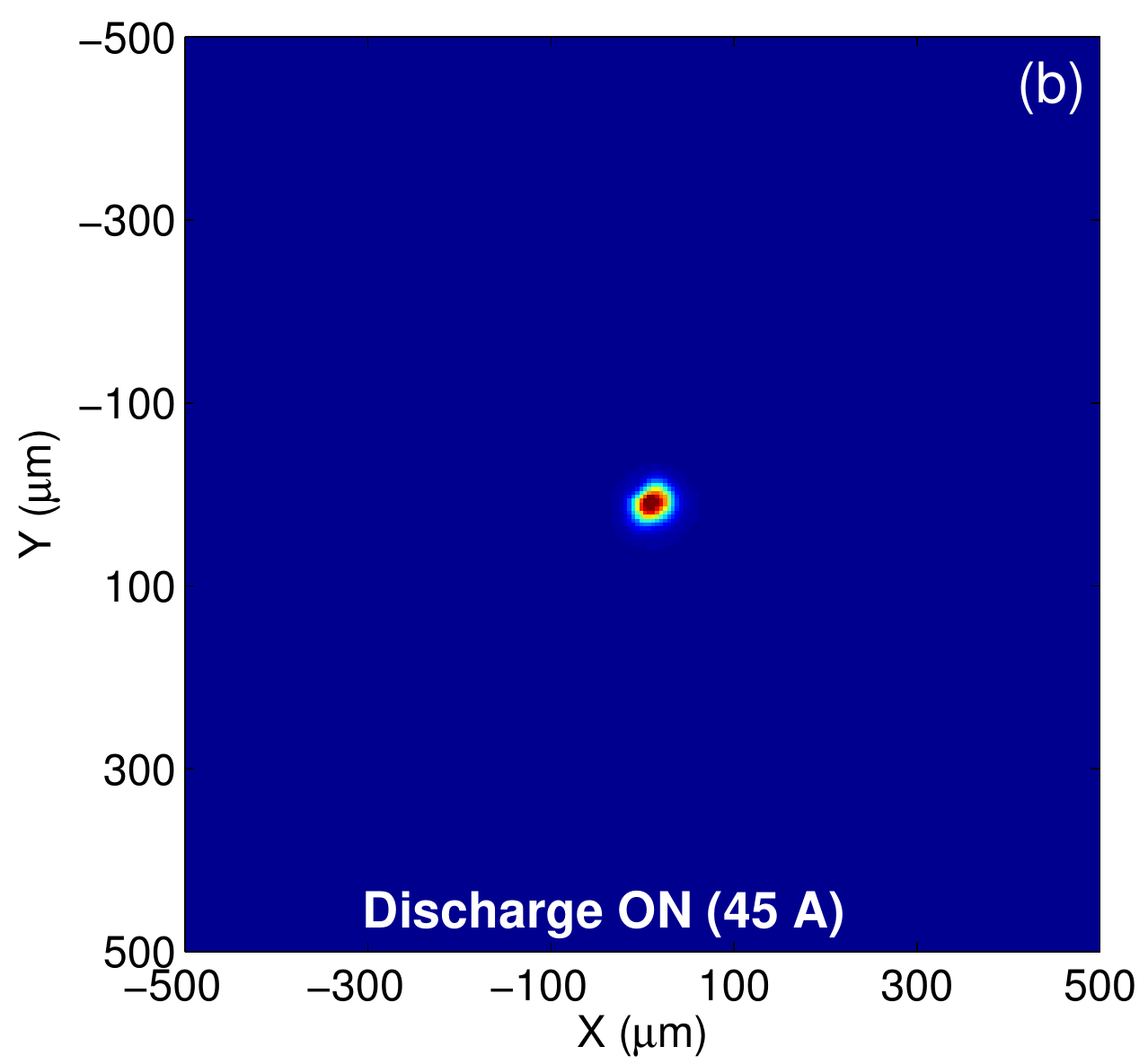}
\put(20,80){\color{white}\textbf{b}}
\end{overpic}
\label{SpotFocused}
}
\subfigure{
\begin{overpic}[width=0.3\textwidth]{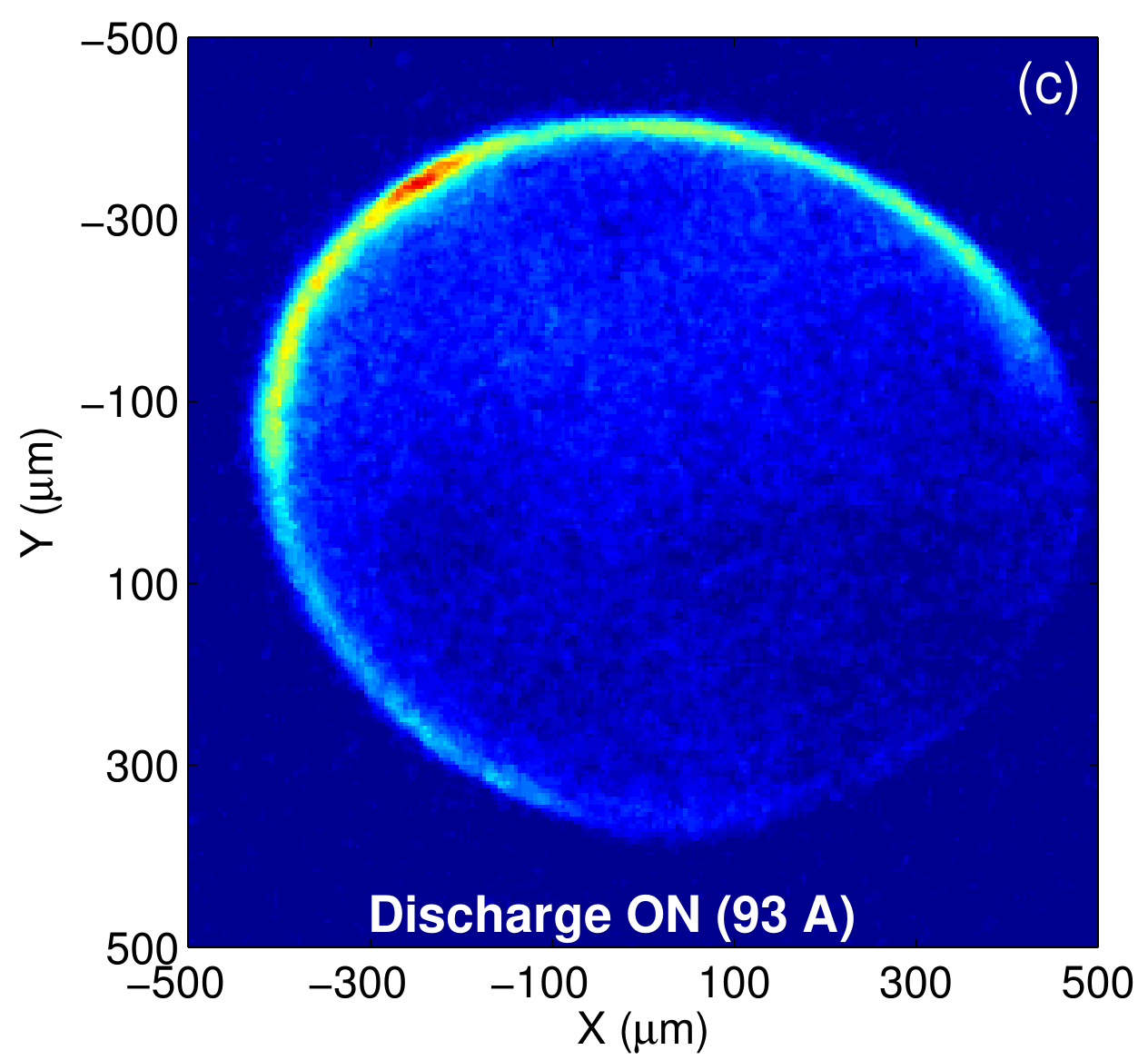}
\put(20,80){\color{white}\textbf{c}}
\end{overpic}
\label{SpotDefocused}
}
\caption{Plasma lens measurements. (a) Beam spot on the first screen downstream the capillary with the discharge off, turned on at 45~A (b) and at 93~A (c).}
\label{plasma_lens}
\end{figure*}

Without any plasma in the capillary, the bunch mean energy is 119.2~MeV with 25~keV energy spread. 
By turning on the HV circuit and synchronizing the passage of the beam through the plasma $\approx 2~\mu s$ after the begin of the discharge\footnote{In order to not have the discharge current overlapped with the beam, thus avoiding any active plasma lensing effect.} ($\approx 10^{15}$~cm$^{-3}$ average background plasma density) the bunch actually acts as a driver losing a small amount of energy, as shown in Fig.~\ref{energy_withplasma}. The mean energy decreases to 119.0~MeV while the energy spread grows up to 47~keV.
These results demonstrate that although the transverse and longitudinal bunch sizes were not optimized (determining a relatively low bunch density), there is still an interaction between the plasma and the bunch that loses part of its energy through it.

With the same capillary setup we have also performed several tests regarding the active plasma lens concept~\cite{pompili2017experimental,lens_alberto}. We measured the full 6D beam phase-space (energy, duration and emittance) analyzing the effects on the beam introduced by the lens. Three consecutive screens downstream the capillary have been employed in order to characterize the beam evolution during its focusing.
Figure~\ref{plasma_lens} shows the transverse squeezing of the beam provided by the active plasma lens for different discharge currents. With the discharge off the spot size is $\sigma_{x,y}\approx 110~\mu m$. When turned on at 45~A we reach the maximum focusing ($\sigma_{x,y}\approx 24~\mu m$)
while for larger values (93~A) the spot size increases to $\sigma_{x,y}\approx 280~\mu m$. 
In addition to this we have found an overall degradation of the beam emittance due to the generation of a nonlinear focusing field. At the capillary entrance the normalized emittance was $1~\mu m$ while in correspondence of the minimum spot size we have measured a larger value ($\approx 3.6~\mu m$). Such a result demonstrates that severe effects are introduced on the beam by the plasma lens with such a capillary configuration.
In this regard, several other configurations will be tested with the aim to obtain the best compromise in terms of emittance preservation and effective focusing.

\section{Conclusions}
The recent experimental activities performed at the SPARC\_LAB test-facility have been presented. Plasma-based focusing has been successfully demonstrated by means of the active plasma lens and the study was completed by a complete characterization of the involved beam dynamics that underlies all the effects introduced on the beam. In view of the PWFA experiment, a detailed description of the experimental apparatus has been provided. It consists in several sub-systems regarding the diagnostics, injection and extraction of the beam. Several preliminary results have been included too.
First tests related to the acceleration of a low charge ($\approx 20$~pC) witness bunch are expected at the beginning of 2018.




\section*{Acknowledgments}
This work has been partially supported by the EU Commission in the Seventh Framework Program, Grant Agreement 312453-EuCARD-2, the European Union Horizon 2020 research and innovation program under Grant Agreement No. 653782 (EuPRAXIA), and the Italian Research Minister in the framework of FIRB - Fondo per gli Investimenti della Ricerca di Base, Project nr. RBFR12NK5K.


\bibliographystyle{elsarticle-num}
\bibliography{biblio}







\end{document}